%
%
\newfam\scrfam
\batchmode\font\tenscr=rsfs10 \errorstopmode
\ifx\tenscr\nullfont
	\message{rsfs script font not available. Replacing with calligraphic.}
\else	\font\sevenscr=rsfs7 
	\font\fivescr=rsfs5 
	\skewchar\tenscr='177 \skewchar\sevenscr='177 \skewchar\fivescr='177
	\textfont\scrfam=\tenscr \scriptfont\scrfam=\sevenscr
	\scriptscriptfont\scrfam=\fivescr
	\def\scr{\fam\scrfam}
	\def\cal{\scr}
\fi
\newfam\msbfam
\batchmode\font\twelvemsb=msbm10 scaled\magstep1 \errorstopmode
\ifx\twelvemsb\nullfont\def\Bbb{\bf}
	\message{Blackboard bold not available. Replacing with boldface.}
\else	\catcode`\@=11
	\font\tenmsb=msbm10 \font\sevenmsb=msbm7 \font\fivemsb=msbm5
	\textfont\msbfam=\tenmsb
	\scriptfont\msbfam=\sevenmsb \scriptscriptfont\msbfam=\fivemsb
	\def\Bbb{\relax\ifmmode\expandafter\Bbb@\else
 		\expandafter\nonmatherr@\expandafter\Bbb\fi}
	\def\Bbb@#1{{\Bbb@@{#1}}}
	\def\Bbb@@#1{\fam\msbfam\relax#1}
	\catcode`\@=\active
\fi
%
%
\font\eightrm=cmr8	\def\xrm{\eightrm}
\font\eightbf=cmbx8	\def\xbf{\eightbf}
\font\eightit=cmti8	\def\xit{\eightit}
\font\eighttt=cmtt8	\def\xtt{\eighttt}
\font\eightcp=cmcsc8

\font\twelverm=cmr12
\font\twelvecp=cmcsc10 scaled\magstep1

%
%
\headline={\ifnum\pageno=1\hfill\else
{\eightcp B.E.W. Nilsson: 
	``A superspace approach to branes and supergravity''}
		\dotfill\folio\fi}
\def\makeheadline{\vbox to 0pt{\vss\noindent\the\headline\break
\hbox to\hsize{\hfill}}
	\vskip2\baselineskip}
%
%
\footline={}
\def\makefootline{\baselineskip=1.6cm\line{\the\footline}}
%
%
\newcount\sectioncount
\sectioncount=0
\def\section#1#2{\global\eqcount=0
	\global\advance\sectioncount by 1
	\vskip2\baselineskip\noindent
	\hbox{\twelvecp\the\sectioncount. #2\hfill}\vskip\baselineskip
	\xdef#1{\the\sectioncount}}
\def\appendix#1{\vskip2\baselineskip\noindent
	\hbox{\twelvecp Appendix: #1\hfill}\vskip\baselineskip}
%
%
\newcount\refcount
\refcount=0
\newwrite\refwrite
\def\ref#1#2{\global\advance\refcount by 1
	\xdef#1{\the\refcount}
	\ifnum\the\refcount=1
	\immediate\openout\refwrite=\jobname.refs
	\fi
	\immediate\write\refwrite
		{\item{[\the\refcount]} #2\hfill\par\vskip-2pt}}
\def\refout{\catcode`\@=11
	\immediate\closeout\refwrite
	\vskip2\baselineskip
	{\noindent\twelvecp References}\hfill\vskip\baselineskip
	\parskip=.875\parskip 
	\baselineskip=.8\baselineskip
	\input\jobname.refs 
	\parskip=8\parskip \divide\parskip by 7
	\baselineskip=1.25\baselineskip 
	\catcode`\@=\active}
%
%
\newcount\eqcount
\eqcount=0
\def\Eqn#1{\global\advance\eqcount by 1
	\xdef#1{\the\sectioncount.\the\eqcount}
		\eqno(\the\sectioncount.\the\eqcount)}
\def\eqn{\global\advance\eqcount by 1
	\eqno(\the\sectioncount.\the\eqcount)}
%
%
\parskip=3.5pt plus .3pt minus .3pt
\baselineskip=12pt plus .1pt minus .05pt
\lineskip=.5pt plus .05pt minus .05pt
\lineskiplimit=.5pt
\abovedisplayskip=10pt plus 4pt minus 2pt
\belowdisplayskip=\abovedisplayskip
\hsize=15cm
\vsize=18.9cm
\hoffset=1cm
\voffset=1.4cm
%
%
\def\/{\over}
\def\*{\partial}
\def\a{\alpha}
\def\b{\beta}

\def\e{\varepsilon}

\def\l{\lambda}

\def\s{\sigma}

\def\G{\Gamma}

\def\.{.\hskip-1pt }

\def\-{\!-\!}
\def\+{\!+\!}

\def\f{{\lower1.5pt\hbox{$\scriptstyle f$}}}
\def\ff{{\lower1pt\hbox{$\scriptstyle f$}}}

%
%
%
%
%
\null\vskip-1cm
\hbox to\hsize{\hfill G\"oteborg-ITP-98-09}
\hbox to\hsize{\hfill\tt hep-th/0007017}
\hbox to\hsize{\hfill January 1998}

\vskip2cm

\vskip1.5cm
\centerline{\twelvecp A superspace approach to branes and supergravity*}

\vskip\parskip
\centerline{\twelvecp}

\vskip2cm
\centerline{\twelverm  Bengt E.W. Nilsson}

\vskip1cm
\centerline{\it Institute for Theoretical Physics}
\centerline{\it Chalmers University of Technology and G\"oteborg University}
\centerline{\it S-412 96 G\"oteborg, Sweden}

\vskip1.5cm
\noindent \underbar{Abstract:} Recent developments in string and M theory 
rely heavily on supersymmetry
suggesting that a revival of superspace techniques in ten and eleven
dimensions may be advantageous. Here we discuss three topics of current
interest where superspace is already playing an important role and where
an improved understanding of superspace  might provide additional insight
into the issues involved.

\vskip1cm
*In Proceedings of the 31st International Symposium Ahrenshoop, September
2-6, 1997, Buckow, Germany. Eds H. Dorn, D. L\"{u}st and G. Weigt (Wiley 1998).
\vfill

\eject

\ref\G{M.B.~Green, {\xtt hep-th/9712195}.}

\ref\HNvP{P.~Howe, H.~Nicolai and A.~Van Proeyen, \xit Phys. Lett. 
\bf 112B \xrm (1982) 446.}

\ref\BNXXXI{B.E.W.~Nilsson, \xit Phys. Lett. \xbf 175B \xrm(1986) 319.}

\ref\HU{P.S.~Howe and A.~Umerski, \xit Phys. Lett. \xbf 177B \xrm (1986) 163.}

\ref\BNXXXIV{B.E.W.~Nilsson and A.~Tollst\'en, \xit Phys. Lett. \xbf 181B
\xrm (1986) 63.}

\ref\CNS{M.~Cederwall, B.E.W.~Nilsson and P.~Sundell, {\xtt hep-th/9712059}.}

\ref\Bandos{I.~Bandos, K.~Lechner, A.~Nurmagambetov, P.~Pasti, D.~Sorokin
 and M.~Tonin,
\xit Phys. Rev. Lett. \xbf 78 \xrm (1997) 4332, ({\xtt hep-th/9701037}).}

\ref\HoweSezgina{P.S.~Howe and E.~Sezgin, \xit Phys. Lett. \xbf B390 
\xrm (1997) 133,  ({\xtt hep-th/9607227}).}

\ref\HoweSezginb{
P.S.~Howe and E.~Sezgin, \xit Phys. Lett. \xbf B394 \xrm (1997) 62 
({\xtt hep-th 9611008}).}

\ref\ACGHN{
T.~Adawi, M.~Cederwall, U.~Gran, M.~Holm and B.E.W.~Nilsson, 
 {\xtt hep-th/9711203}.}

\ref\BDHS{
K.~Bautier, S.~Deser, M.~Henneaux and D.~Seminara, \xit Phys. Lett.
 \xbf B406 \xrm (1997) 49 ({\xtt hep-th/9704131}).}

\ref\Howe{
P.S.~Howe, {\xtt hep-th/9707184}.}

\ref\BNVII{
B.E.W.~Nilsson, \xit Nucl. Phys \xbf B188 \xrm (1981) 176.}

\ref\BNXXVIII{
B.E.W.~Nilsson and A.~Tollst\'en, \xit Phys. Lett \xbf 171B \xrm (1986) 212.}

\ref\BPT{
L.~Bonora, P.~Pasti and M.~Tonin, \xit Phys. Lett \xbf B188 \xrm (1987) 335.}

\ref\DSW{
M.~de Roo, H.~Suelmann and A.~Wiedemann, \xit Phys. Lett. \xbf B280 
\xrm (1992) 39.}

\ref\BSTa{
E.~Bergshoeff, E.~Sezgin and P.K.~Townsend,
\xit Ann. of Phys \xbf 18 \xrm (1987) 330.}

\ref\BNLIX{
M.~Cederwall, A.~von Gussich, A.~Mikovic, B.E.W.~Nilsson and 
A.~Westerberg, \xit Phys. Lett. \xbf 390B \xrm (1997) 148 
({\xtt hep-th/9606173}).}

\ref\BNLX{
M.~Cederwall, A.~von Gussich, B.E.W.~Nilsson and 
A.~Westerberg, \xit Nucl. Phys. \xbf B490 \xrm (1997) 163 
({\xtt hep-th/9610148}).}

\ref\BNLXI{
M.~Cederwall, A.~von Gussich, B.E.W.~Nilsson, P.~Sundell and 
A.~Westerberg, \xit Nucl. Phys. \xbf B490 \xrm (1997) 179 
({\xtt hep-th/9611159}).}

\ref\CT{
M.~Cederwall and P.K.~Townsend, {\xtt hep-th/9709002}.}

\ref\Witten{
E.~Witten, {\xtt hep-th/9610234}.}

\ref\HSW{
P.S.~Howe, E.~Sezgin and P.~West, \xit Phys. Lett. \xbf B399 
\xrm (1997) 49 ({\xtt hep-th/9702008}).}

\ref\BG{
J.~Bagger and A.~Galperin, \xit Phys. Rev. \xbf D55 \xrm (1997) 1091 
({\xtt hep-th/9608177}).}

\section\introduction{Introduction}

It is well-known that besides the string (the $p=1$ brane) 
itself also branes
of other dimensionality ($d=p+1$) play a central role in  
the non-perturbative structure of the theory. Using Dp-branes
 and T5-branes (having respectively vector and second rank
antisymmetric tensor potentials on the world surface) 
non-perturbative relations 
between all string theories can be established as well
as a connection with 11d supergravity. This hints
at the existence of an underlying more profound formulation of the whole
theory. Some aspects of this so called M theory are captured by 
M(atrix) theory which has its origin in D-particle (D0-brane) 
physics, but have many 
features in common with the 11d membrane (see e.g. the talk by 
B. de Wit).

The three topics discussed below are related to the role 
superspace is playing in this context. 
We start by discussing one of the direct
implications of formulating the fundamental theory in terms of
M(atrix) theory, namely the presence of 
higher order corrections (e.g. $R^n$)
in 11d supergravity (for a recent review see [\G]).
 Here we will have reason to recall
certain facts in 10d supergravity established in the
1980's [\HNvP, \BNXXXI, \HU, \BNXXXIV].
Then we review some recent results concerning the realization of
 $\kappa$-symmetry
for various branes focusing on 
 the 11d membrane ($p=2$),
the D3-brane in 10d
type IIB supergravity, and the T5-brane
 in 11d superspace. 
In the latter case we present a new action [\CNS]
 which avoids some of the problematic features of the lagrangian
constructed previously in [\Bandos].
As a final topic we consider the 
generalization of this setup, based on bosonic world sheets embedded into
target superspaces, to a situation where both target and the embedded
world surface are superspaces [\HoweSezgina, \HoweSezginb, \ACGHN].
 In [\ACGHN] the relation between non-linearities of
the tensor dynamics on the world sheet and non-linearly realized 
(super)symmetries is analyzed. This paper also contains a
 detailed account of general embeddings plus
some new results for T5-branes in 7d.

\section\corrections{Superspace and higher order corrections in 11d}

One property of M theory that has been highlighted by the recent developments
in M(atrix) theory is the higher order corrections in terms of e.g. the
curvature tensor that must occur in the low energy supergravity lagrangian
in 11d. Since 11d supersymmetry is tremendously
restrictive (there is only one multiplet whose spin content does not
exceed spin two, a cosmological term is not possible [\BDHS]
, etc) but also very 
messy to deal with in terms of component fields, it might be
worth the effort to develop further the  superspace techniques that were
introduced in the 1970's. 
The heart of the matter is the question of 
how to realize supersymmetry off-shell. This 
is fairly well understood in the case of $N=1$ supersymmetry in 10d 
[\HNvP, \BNXXXI, \HU, \BNXXXIV],
a subject that we will therefore have reason to come back to.

The field content of 11d supergravity is
$
{e_m}^a,\,\,\psi_m^{\a},\,\,c_{mnp}
$
where $m,n,..$ are 11d vector indices and $\a=1,..,32$ is a Majorana 
spinor index. These fields describe 128 bosonic and 128 fermionic 
degrees of freedom. The field 
equations are obtained from the corresponding superfields
and their super-Bianchi identities. To this end we combine 
${e_m}^a$ and $\psi_m^{\a}$
into the supervielbein
$
{E_M}^A
$
where the superworld index $M=(m,\mu)$ and the supertangent space index 
$A=(a,\a)$. Normally one introduces also a superfield  $C_{MNP}$
with $c_{mnp}$ as its first component.
However, it was recently clarified [\Howe] 
that this is not necessary, and 
we will refrain from doing so. Then the field strengths
are just the supercurvature ${R_A}^B$ and the supertorsion 
$T^A=DE^A=dE^A-E^B{{\omega}_B}^A$. Their Bianchi identities read
$
DT^A={E^B}{R_B}^A$ and $D{R_A}^B=0
$
but in fact it is only necessary to consider the first identity
 since the second
one is automatically satisfied. The meaning of 
 ``solving the
Bianchi identities'' is as follows. By subjecting
the torsion components (${T_{\a \b}}^c$, etc) to certain 
constraints the Bianchi
identities cease to be identities and become
 equivalent to the field equations.
One should remember however that a given set of gauge fields can often 
be given a 
variety of different kinds of dynamics, related either to different sets of 
constraints or to differences in the Bianchi identities.
As shown recently by Howe [\Howe] (on-shell)
 11d supergravity is obtained from the $single$ 
constraint ($\Gamma^c$ is an 11d Dirac matrix)
$$
{T_{\a \b}}^c = 2i({\Gamma^c})_{\a \b}\Eqn\t
$$
which is invariant under 
super-Weyl rescalings. 
Furthermore, no off-shell formulation of this theory is known.

This situation should be compared to what is known for $N=1$ supergravity
in 10d. When the Bianchi identities are solved on-shell [\BNVII] 
one finds that
all physical fields appear at different $\theta$ levels in a scalar 
superfield  denoted $\Phi(Z)$, where $Z=(x,\theta)$.
Note that 10d supergravity
contains a scalar and a spinor, apart from 
$
{e_m}^a,\,\,\psi_m^{\a},\,\,B_{mn}
$. In contrast to 11d, one must 
in this case constrain several
torsion components.
This theory can be coupled to superYang-Mills using the same torsion 
constraints [\BNXXVIII] and the 
 Bianchi identity  $DH=trF^2$ 
for the three-form $H=dB$. In fact, as proven in [\BPT]
even the $R^2$ term needed for anomaly cancellation can be 
dealt with without altering the constraints. 

However, also $R^4$ and higher 
terms are present in the field theory of the low energy 10d superstring.
 In general such terms 
cannot be incorporated into the superspace Bianchi identities 
if the on-shell constraints 
 are used. Fortunately, in this case it is known how
to proceed since the off-shell 
 field content is known. It consists of a superconformal gravity
 multiplet and 
an unconstrained scalar auxiliary superfield $w$ [\HNvP].
To account for this new superfield the constraints must be modified to
[\BNXXXI,\HU]
$$
{T_{\a \b}}^c=2i(\Gamma^c)_{\a \b} + 
2i(\Gamma^{c_1...c_5})_{\a \b}{X^c}_{c_1...c_5}\Eqn\tx
$$
where $X$ is in the representation $1050^+$ of $so(1,9)$
 appearing at level $\theta^4$ in $w$.

All higher order corrections, like $R^4$, that can 
occur must be compatible with supersymmetry and fit
somewhere in the solution of the Bianchi identities that follow from the
off-shell constraints above. E.g. the $R^4$ term
 related by supersymmetry (see [\DSW]) to the anomaly
term $BX_8$ can be added as follows [\BNXXXIV]:
$$
S^{D=10,N=1}=-{1\/{{({\kappa}_{10})}^2}}\int d^{10}xd^{16}\theta E\Phi(w+c)
\Eqn\s
$$
where $c$ is a constant proportional to $\zeta(3)$, and where the $w$ term
is the kinetic one and the $c$ term is the supersymmetrization of $R^4$.
$E$ is the superspace measure. 

Turning to 11d the situation changes dramatically since 
it is not known how to solve the Bianchi identities off-shell or 
how to write down an
off-shell action in components. This makes it very hard to address 
questions in
11d supergravity concerning 
the higher order corrections.
We will here introduce the equivalent of $w$ in 10d
into the 11d supertorsion by means of the relaxed constraint
$$
{T_{\a\b}}^c=2i{\Gamma^c}_{\a\b}+2i{\Gamma^{d_1d_2}}_{\a\b}{X^c}_{d_1d_2}
+2i{\Gamma^{d_1...d_5}}_{\a\b}{X^c}_{d_1...d_5}\Eqn\txx
$$
where the tensors in the last two terms are in the representations 
429 and 4290 which
appear at level $\theta^4$ in an unconstrained 11d scalar superfield.
 A preliminary analyzes
of the Bianchi identities indicates an ``off-shell situation''
where new terms appear in the torsion  which could
account for the higher order corrections. In particular
 terms generated by anomalies and the presence of branes should be
investigated.
 Note that
the T5-brane produces in the supersymmetry algebra a 
five-form central charge, a fact that should be compared to
 the extra terms in the torsion. 

Further studies will hopefully tell if these techniques can be utilized
in the endeavour to extract an 11d supergravity theory from M(atrix) theory 
or perhaps directly from the 11d branes.

\section\bosonicbranes{$\kappa$ symmetric branes as bosonic surfaces}

The relevant branes in 11d are the membrane 
and the T5-brane, while in 10d type II theories
there are also Dp-branes intermediate between p-branes and T-branes. 
As we will see below, for branes
with vector or tensor fields propagating on the world
surface $\kappa$-symmetry is technically more complicated than for
ordinary p-branes. 

Let us as an example of an ordinary p-brane consider the 11d membrane 
[\BSTa]. After the elimination  (see [\BNLIX]) 
of the independent world sheet
metric by means of its algebraic field equation the action reads:
$$
S_3=\int d^3 \xi [-\sqrt{-detg}-{\e}^{ijk}B_{ijk}]\Eqn\sm
$$
where ${g}_{ij}$ is the pull-back of the target space metric, i.e.
${g}_{ij}={\Pi}^a_i{\Pi}^b_j{\eta}_{ab}$, and 
 the ${\xi}^i$'s are three bosonic coordinates on the world sheet. 
The background superfields ${E_M}^A$ and $B_{MNP}$ 
of 11d
supergravity enter via the pull-backs
 ${\Pi}^A_i=\partial_iZ^M{E_M}^A$ and 
$B_{ijk}={\Pi}^A_i{\Pi}^B_j{\Pi}^C_kB_{CBA}$. 
In order for this action to
be supersymmetric the number of bosonic (here 11-3=8) and
 fermionic ($32\times{1\/2}$)
world sheet on-shell degrees of freedom must
match. This requires the presence
 of the local fermionic
$\kappa$-symmetry giving another factor of ${1\/2}$ in the fermionic count.
 Its existence relies on the possibility to construct
a projection operator ${1\/2} (1+\Gamma)$ with $\Gamma^2=1$. Here 
$\Gamma={1\/6{\sqrt{-g}}}{\epsilon}^{ijk}{\Gamma}_{ijk}$
where ${\Gamma}_{ijk}$ is the pull-back
of $\Gamma_{abc}=\Gamma_{[a}\Gamma_b\Gamma_{c]}$.

This structure can be found also in the case of Dp-branes [\BNLX,\BNLXI],
 but is now 
more involved due to the presence of the field strength $F_{ij}$.
E.g. the D3-brane in the 10d type IIB theory has an action that reads
$$
S_{D3}=-\int d^4{\xi} \sqrt{-det(g+e^{-{\phi\/2}}{\cal F})}+\int e^{\cal F}C
\Eqn\sd
$$
In this case there are, apart from the dilaton $\phi$, 
two kinds of background potentials $B$ and $C$, coming from
the NS-NS
 and the R-R sector, respectively.
 In $S_{D3}$, ${\cal F}_{ij}=F_{ij}-B_{ij}$ where 
 $B_{ij}$ is the pullback of $B_{MN}$ in the 10d IIB
target space theory. The last term in the action is constructed as a formal sum
of forms of different rank and the integral is supposed to pick up only
the four-form in this case. That the action is $\kappa$-symmetric can then be
shown using the $\Gamma$ matrix $(\Gamma^2=1)$ [\BNLX, \BNLXI]
$$
\Gamma={{\epsilon}^{ijkl}\/{\sqrt{-det(g+{\cal F})}}}({1\/24}\Gamma_{ijkl}I
-{1\/4}{\cal F}_{ij}{\Gamma}_{kl}J+{1\/8}{\cal F}_{ij}{\cal F}_{kl})\Eqn\G
$$ 
 The $SL(2;Z)$ symmetry of the IIB theory mixes the $B$ 
and $C$ potentials and indeed a more symmetric version of the action
exists [\CT]. In that version all background 
potentials have associated world sheet field strengths to which they couple
as $F-B$.

The third case to be discussed here is the T5-brane in 11d. This brane has 
an additional complication in that the three-form field strength is self-dual.
Finding a covariant action for such a field has been a long-standing
problem. A first attempt at a solution was given recently by Bandos
et al in [\Bandos]) and involves an auxiliary scalar field $a$
entering the lagrangian through a factor ${1\/{{(\partial a)}^2}}$. Objections
against using such an action in quantum calculations have been formulated (see 
also [\Witten]). Another action that does not make use of
such a scalar was subsequently presented in [\CNS]. Although
 the problems associated with the scalar are gone, the 
other objections
in [\Witten] probably remain. Nevertheless, since this action 
exhibits some new features 
it might be of some interest. The action is (${\cal F}$ is an independent 
six-form field strength)
$$
S_{T5}=\int d^6\xi \sqrt{-g}\l (1+{1\/12}F_{ijk}F^{ijk}-{1\/24}k_{ij}k^{ij}
+{1\/72}(trk)^2-(*{\cal F})^2)\Eqn\st
$$
where $k_{ij}={1\/2}{F_i}^{jk}F_{jkl}$. Note that, as explained in 
[\CNS],
the self-duality relation 
$$
-(*{\cal F})*F_{ijk}=F_{ijk}-{1\/2}{k_{[i}}^lF_{jk]l}+{1\/6}(trk)F_{ijk}
\Eqn\fsd
$$
does not arise
as a field equation but as a result of demanding $\kappa$-symmetry, 
and must not be inserted into the action.

\section\superbranes{Superworld sheets  in target superspace}

It is possible to reformulate the brane dynamics, following from 
actions of the kinds described in the previous section, in terms of
world sheet superfields. This can be done by embedding superworld sheets into
target superspace. Besides making the world sheet supersymmetry
manifest this has the further advantages of explaining the origin of
$\kappa$-symmetry and the projection matrix [\HSW] as well as of providing
the connection between superembeddings on one hand, and Goldstone fermions 
and non-linearly
realized supersymmetries on the supersheets on the other hand 
[\BG, \ACGHN].

In [\BG] Bagger and Galperin showed how the 4d Born-Infeld action for
an abelian gauge field can be obtained by embedding a (4d,N=1) superspace
into a (4d,N=2) one. The broken fermionic translations turn into non-linear
supersymmetry transformations on the Goldstone fermions arising from half of
the fermionic coordinates that are turned into dependent
variables. The non-linearities of the Born-Infeld action are then seen to be
a consequence of demanding consistency with the extra non-linear 
supersymmetries. In [\ACGHN] this programme was taken over to
the T5-brane in 7d superspace. This led to an equation for
 one of the supertorsion components whose rather 
complicated solution indicates that this way of analyzing
this system is not the most efficient one. Fortunately once it is realized
that this form of the torsion equation can also be obtained in the much
more general formalism known as the embedding formalism developed in
[\HoweSezgina, \HoweSezginb] 
 these problems
can be circumvented. 

The central equation is the torsion pullback equation
 (\HoweSezgina,\HoweSezginb,\ACGHN)
$$
{\cal D}_A{{\cal E}_B}^{\underline C}-
{(-1)}^{AB}{\cal D}_B{{\cal E}_A}^{\underline C}+
{T_{AB}}^C{{\cal E}_C}^{\underline C}=
(-1)^{A(A+{\underline B})}
{{\cal E}_B}^{\underline B}{{\cal E}_A}^{\underline A}
{T_{{\underline A}{\underline B}}}^{\underline C}\Eqn\e
$$  
where underlined indices refer to target superspace while the other indices
are connected with the superworld sheet. As shown in [\HSW, \ACGHN]
inserting constraints on the torsion components turns this equation into
the equations of motion for the world sheet fields. In particular, the 
highly non-linear dynamics of the T5-branes in 11d [\HSW] and in 7d
[\ACGHN] can be obtained this way. 

{\bf Acknowledgement}: I wish to thank my coauthors on refs. 
[\BNXXXIV, \CNS, \ACGHN, \BNXXVIII, \BNLX, \BNLXI]
for very nice and fruitful collaborations.

\xrm\refout 

\end